\documentclass{emulateapj}
\slugcomment{Accepted to ApJL}

\begin{document}
\title{Linearly polarized X-ray flares following short Gamma-ray Bursts}
%\footnote{}

\author{Y. Z. Fan$^{1,2,3}$, Bing Zhang$^{3}$, Daniel Proga$^{3}$}
\affil{$^1$ Purple Mountain Observatory, Chinese Academy of
Science, Nanjing 210008, China.\\
$^2$ National Astronomical Observatories, Chinese Academy of
Sciences, Beijing, 100012, China.\\
$^3$ Department of Physics, University of Nevada, Las Vegas, NV
89154, USA.\\ }

\begin{abstract}
Soft X-ray flares were detected to follow the short-duration
gamma-ray burst GRB 050724. The temporal properties of the flares
suggest that they are likely due to the late time activity of the
central engine.  We argue that if short GRBs are generated through
compact star mergers, as is supported by the recent observations, the
jet powering the late X-ray flares must be launched via magnetic
processes rather than via neutrino-antineutrino annihilations.  As
a result, the X-ray flares following short GRBs are expected to be
linearly polarized. The argument may also apply to the X-ray
flares following long GRBs. Future observations with the upcoming
X-ray polarimeters will test this prediction.
\end{abstract}


%%%%%%%%%%%%%%%%%%%%%%%%%%%%%%%%%
\section{Introduction}

Recently major breakthroughs were made to understand short-duration
gamma-ray bursts (GRBs). The {\em Swift} satellite made the first
localization of a short-hard burst, GRB 050509B (Gehrels et
al. 2005), which is proposed to be associated with a luminous
elliptical galaxy at $z=0.225$ (Gehrels et al. 2005; Bloom et
al. 2005). Deep search of an underlying supernova has result in null
results (Hjorth et al. 2005a). Two months later, HETE-2
localized the second short burst, GRB 050709 (Villasenor et al. 2005),
whose optical and X-ray afterglows were detected (Fox et al. 2005;
Villasenor et al. 2005). The long-wavelength afterglow counterpart is
in a star forming galaxy at redshift $z=0.16$, but late time
monitoring again places tight constraints on the existence of a
possible underlying
supernova (Covino et al. 2005; Fox et al. 2005; Hjorth et al. 2005b).
Shortly after, {\em Swift} localized another one, GRB 050724
(Barthelmy et al. 2005). Being more luminous, its X-ray, radio,
optical and infrared afterglows were all well detected (Romano et
al. 2005; Cameron \& Frail 2005; Gal-Yam et al. 2005; D'Avanzo et
al. 2005).  This burst is within an elliptical galaxy at redshift
$z=0.257$ (Prochaska et al. 2005; Barthelmy et al. 2005). The spectral
information
%(e.g. the lack of detected Balmer H$\beta$ absorption,
%Berger et al. 2005)
indicates that the host is again an early type
galaxy, with a stellar population older than $\sim 1$ Gyr. The overall
star formation rate is estimated to be lower than $0.03 M_\odot~{\rm
yr}^{-1}$ (Berger et al. 2005; Barthelmy et al. 2005).

All the pieces of evidence seem to support the suggestion
that short GRBs are produced from mergers of two compact objects
rather than from collapsar-related events (see Fan et al. 2005
for a general discussion on various models for short GRBs).
The commonly discussed scenarios (e.g. Eichler et al. 1989; Paczy\'nski
1991; Narayan et al. 1992; Fryer et al. 1999a) include
double neutron star (NS-NS) mergers, mergers between a neutron star
and a preexisting black hole of several solar masses (NS-BH), and
mergers between a white dwarf and a black hole (WD-BH). The WD-BH
scenario is less favored at least for GRBs 050509B and 050724, since
they are expected to occur in star forming regions and can not sit in
the outskirt of the host galaxy. Also the disk may be too large to
efficiently launch a relativistic jet to power a GRB (Narayan et
al. 2001). On the other
hand, NS-NS and NS-BH mergers are expected to occur in early type
galaxies with an old stellar population, and they are expected to
occur in regions with a large offset from the host galaxy center
(sometimes even at the outskirt of the host galaxy), due
to the asymmetric kicks during the formation of the NSs
(e.g. Bloom et al. 1999). Numerical simulations suggest that the
typical coalescence time scale for NS-NS mergers (e.g., Eichler
et al. 1989; Narayan et al. 1992; Davies et
al. 1994; Ruffert \& Janka 1998; Rosswog et al. 2000; Narayan et
al. 2001; Rosswog et al. 2003; Aloy et al. 2005) is short,
e.g. $t_{acc} \sim 2.76\times 10^{-2}\alpha_{-1}^{-6/5}~{\rm s}$,
where $\alpha \sim 0.1$ is the viscosity parameter (Narayan et
al. 2001; Popham et al. 1999). A similar time scale is also derived
for NS-BH mergers although arguments disfavoring such a model have
been raised recently (e.g. Miller 2005; Rosswog 2005).
Nonetheless, the host galaxy identifications and the observed short
durations of the hard spike in these GRBs are generally consistent
with the merger models, especially the one involving NS-NS mergers.

Here we stress an important new phenomenon, i.e. the soft X-ray
flares (lasting for a few hundred seconds or even longer) detected
in GRB 050724. We argue that the flares are ejected directly from
the central engine and should be of magnetic-origin. The emission
in the flares should therefore be linearly polarized.

\section{X-ray Flares following Short-Hard GRB 050724}

The prompt emission (15-25 keV) of this GRB 050724
contains two emission components, i.e. a short-hard pulse
followed by a long-soft emission component lasting longer that 100
s (see Fig.1b of Barthelmy et al. 2005). The long, soft emission
component is spiky. This is confirmed by the XRT
observations starting from 79 s after the trigger which overlaps
the BAT observation of the soft component.  The early XRT
lightcurve initially shows a steep decay with a slope $\sim -2$.
This flare-like event is then followed by a very rapid decay after
$\sim 100$ s (the index of the power-law decay $\sim -10$).
Around 200-300 s, a second, less-energetic flare emerges. The
X-ray flux drops rapidly again (with index -7) between 300 s to
400 s and then flattens (see Fig.3 of Barthelmy et al. 2005).

Previously the temporal variabilities in some GRB afterglows have been
interpreted as due to refreshed external shocks (e.g. Granot et
al. 2003 for GRB 030329). In such a scenario, the lightcurve after
each ``refresh'' is boosted and the lightcurve typically shows a
``step-like'' behavior (e.g. Zhang \& M\'esz\'aros 2002a). The decay
slope in any external shock model can not be steeper than $-(2+\beta)$
(where $\beta$ is the spectral index of the emission), a limit set by
the so-called ``curvature effect'' (Kumar \& Panaitescu 2000; see also
Fan \& Wei 2005, Panaitescu et al. 2005 for derivations). The observed
very steep decays (with indices -10 and -7) following the two flares
are therefore not consistent with such an interpretation.

Alternatively, the flares may be the consequence of the late
central engine activity, as has been extensively discussed
recently (e.g. Fan \& Wei 2005; Burrows et al. 2005; Zhang et al.
2005). Within this scenario, if the central engine operates in
well-separated episodes, a steep decay can be expected to follow
the end of each episode. After the central engine turns off, the
tail emission should decline as ${\sum\limits_{i}} F_{\nu_{\rm
x},i} [(t-t_{{\rm eje},i})/\delta t_{i}]^{\rm -(2+\beta_i)}$,
where $i$ represents the $i$th pulse, $F_{\nu_{\rm x},i}$,
$t_{{\rm eje},i}$, and $t_{i}$ are the peak flux, the ejection
time and the variability timescale of the $i$th pulse,
respectively. By choosing the trigger time as the zero point, the
decay slope after a late-time flare could be in principle, (much)
steeper than $-(2+\beta)$ (Fan \& Wei 2005; Zhang et al. 2005).

So the ``X-ray flares" following GRB 050724 is likely produced by
the operation of the central engine (see also Barthelmy et al.
2005; Zhang et al. 2005). However, in the standard compact object
merger scenarios, it is a big challenge to prolong the accretion
episode to be as long as a few hundred seconds. One possible model
mentioned in Barthelmy et al. (2005) invokes a BH-NS merger system
in which the NS may be partially disrupted in the initial
collapse. The X-ray flares are produced by the accretion of these
late clumps not accreted during the prompt emission epoch.
Alternatively, the extended central engine activity may be the
result of an accretion flow modulated by the ``magnetic-barrier''
and gravity (D. Proga et al. 2005, in preparation). MHD
simulations show that accretion can be quenched by the strong
magnetic field that forms a magnetized polar cylinder (magnetic
barrier) around the black hole (e.g., Proga \& Begelman 2003).
Such a barrier would halt the accretion flow intermittently (see
Figs.6 \& 8 in Proga \& Begelman 2003), resulting in an episodic
accretion rate (see Fig.3 of Proga \& Begelman 2003). This
potentially gives a natural mechanism for flaring variability in
the magnetic-origin models. Detailed numerical simulations are
desirable to validate these suggestions. Here instead of proposing
such a mechanism, we simply assume the existence of such a
long-term central engine and turn to investigate the possible
energy extraction mechanism that powers the X-ray flares.

\section{Constraints on the Energy Extraction Mechanism}\label{sec:Extraction}

In the compact object merger scenarios (e.g. NS-NS and NS-BH mergers),
the total mass available for the accretion is $\sim
0.1-1.0~{M_\odot}$. The X-ray flares detected
in GRB 050724 lasted $\sim 100$ s (which is also the time scale of the
central engine). Therefore, even if we assume that most of the
mass is accreted during the X-ray flare phase rather than during the
short, hard spike phase,
the time averaged accretion rate is at most $\sim 0.001-0.01 ~{\rm
M_\odot~s^{-1}}$. This fact alone poses important constraints on the
energy extraction mechanism near the central engine.

Two popular energy extraction mechanisms have been discussed in the
GRB central engine models. Here we discuss them in turn.

{\em Neutrino mechanism}. The first mechanism commonly discussed
invokes neutrino annihilation ($\nu\bar{\nu}\rightarrow e^+e^-$,
e.g. Ruffert \& Janka 1998). The
fireball (jet) luminosity driven by this mechanism very
sensitively depends on the mass accretion rate, since the neutrino
emission sensitively depends on the density and the temperature of the
torus. For accretion rates ($\dot{M}$) between 0.01 and 0.1
$M_\odot~{\rm s^{-1}}$, the $\nu \bar{\nu}$ annihilation luminosity
could be well fitted by (Popham et al. 1999; Fryer et al. 1999b;
Janiuk et al. 2004)
\begin{equation}
\log L_{\nu \bar{\nu}}({\rm ergs~s^{-1}})\approx 43.6+4.89
\log({\dot{M} \over 0.01 M_\odot ~{\rm s^{-1}}})+3.4a,
\label{Eq:Popham99b}
\end{equation}
where $a=J_{\rm BH}c/GM_{\rm BH}^2$ is the spin parameter of the
central BH, $J_{\rm BH}$ and $M_{\rm BH}$ are the angular momentum
and the mass of the central black hole. If one takes the typically
value of $a\sim 0.5$ (Fryer et al. 1999b), the jet luminosity
powered by neutrino annihilation is $L_{\nu \bar{\nu}}<
10^{45}~{\rm ergs~s^{-1}}$. For GRB 050724, the time averaged
isotropic luminosity of the X-ray flare component is $L_{\rm X}
\approx 10^{48}~{\rm ergs~cm^{-2}}$. Since only a fraction of
$L_{\nu\bar{\nu}}$ can be converted into the observed X-ray
emission, the $\nu \bar{\nu}$ annihilation mechanism cannot
provide enough energy to power the X-ray flares detected in GRB
050724, unless the outflow is collimated into a very narrow jet
with a solid angle $\Omega < 0.001$, or with a typical jet
half-opening angle $\theta_j < 0.03$ rad. Without a proper
collimation agent (e.g. a stellar mantle as in the collapsar
scenario or a magnetically driven wind from the disk), the outflow
resulting from a compact object merger is expected to be only
mildly collimated (cf. Guetta \& Piran 2005). This viewpoint is
also supported by the observations of GRB 050709 and GRB 050724,
which inferred $\theta_j$ being as large as 0.3 rad and 0.15 rad,
respectively (Villasenor et al. 2005; Berger et al. 2005). We
therefore conclude that the neutrino mechanism is insufficient to
power the X-ray flares.

{\em Magnetic mechanism}.
Alternatively, a relativistic jet could be
launched from a black hole - torus system through MHD processes.
For example, a MHD numerical simulation for $\dot M \sim 1
{\rm M_\odot /s}$ (Proga et al. 2003) suggests that
the efficiency to convert the accretion luminosity to a
Poynting-flux-dominated outflow luminosity is about a factor of $\sim
10^{-4}-10^{-3}$ (see also Mizuno et al. 2004).
Although no specific simulation has
been carried out for the parameter range $\dot M \sim (0.01-0.001)~{\rm
M_\odot /s}$, a natural expectation is that the efficiency should not
sensitively depend on the accretion rate, because both accretion and
jet formation depend on the same agent, i.e. the magnetic
fields in the accretion flow, and because there is no strong
dependence on the density and temperature in the torus as has been
in the case of
neutrino generation. If we still adopt an efficiency of $\sim 10^{-4}
- 10^{-3}$, the expected jet luminosity should be $\sim
10^{47}-10^{48} ~{\rm ergs~s^{-1}}$, adequate to interpret the
observed luminosities of the X-ray flares even if a very moderate
beaming factor is involved.

Another energy source in the central engine would be
the spin energy of the black hole, which might be tapped by magnetic
fields through the Blandford-Znajek (1977) mechanism
(e.g. M\'esz\'aros \& Rees 1997).
The jet luminosity could be estimated as
$L_{\rm BZ}\approx 2.5\times 10^{47}({a/ 0.5})^2({B /
10^{14}{\rm G}})^2~{\rm ergs~s^{-1}}$,
where $B$ is the magnetic field at the central engine. This power is
also adequate to power the X-ray flares as long as the black hole spin
energy is essentially not tapped during the prompt emission phase. In
such a case, the jet is also Poynting-flux dominated.

In a Poynting-flux-dominated flow, the observed X-ray flare
emission could be due to dissipation of the magnetic fields. By
comparing with the pair density ($\propto r^{-2}$, $r$ is the
radial distance from the central source) and the density required
for co-rotation ($\propto r^{-1}$ beyond the light cylinder of the
compact object), one can estimate the radius at which the MHD
condition breaks down, which reads (e.g. Zhang \& M\'esz\'aros
2002b)
\begin{equation}
r_{\rm MHD} \sim (2\times 10^{15}) L_{48}^{1/2}
\sigma_1^{-1}t_{v,m,-3} \Gamma_2^{-1} ~{\rm cm},
\end{equation}
where $\sigma$ is the ratio of the magnetic energy flux to the
particle energy flux, $\Gamma$ is the bulk Lorentz factor of the
outflow, $t_{v,m}$ is the minimum variability timescale of the
central engine. Beyond this radius, significant magnetic
dissipation processes are expected to happen (e.g. Usov 1994)
which convert energy into radiation. At $r_{\rm MHD}$, the
comoving magnetic fields $B_{\rm MHD}$ can be estimated as (e.g.
Eq.[13] of Zhang \& M\'esz\'aros 2002b)
\begin{equation}
B_{\rm MHD}\sim 50~\sigma_1 t_{v,m,-3}^{-1}~{\rm Gauss}.
\end{equation}
When magnetic dissipation occurs, a fraction $\epsilon_{rm e}$ of
the dissipated comoving magnetic energy would be eventually
converted to the comoving kinetic energy of the electrons.
Electrons may be linearly accelerated in the electric fields, and
we assume that the accelerated electrons have a single power-law
distribution $dn/d\gamma_{\rm e} \propto \gamma_{\rm e}^{\rm -p}$
for $\gamma_{\rm e}>\gamma_{\rm e,m}$, where $\gamma_{\rm e,m}$
can be estimated as
\begin{equation}
\gamma_{\rm e,m}\sim 2.1\times 10^3~\sigma_1 C_p,
\end{equation}
and $C_p \equiv (\epsilon_e/0.5)[13(p-2)]/[3(p-1)]$. At
$r_{\rm MHD}$, the corresponding synchrotron radiation frequency
is
\begin{equation}
\nu_{\rm m,MHD}\sim 6\times 10^{16}~\sigma_1^3 C_p^2 \Gamma_2
t_{v,m,-3} (1+z)^{-1}~{\rm Hz}.
\end{equation}
The cooling Lorentz factor can be estimated by $\gamma_{\rm
e,c}\sim 4.5\times 10^{19}\Gamma /(r_{\rm MHD} B^2) \sim 1000$ for
typical parameters taken here. This is comparable to $\gamma_{\rm
e,m}$. As a result the bulk of the energy of the accelerated
electrons is radiated in the soft X-ray band. If this X-ray
component due to central engine activity dominates over the
forward shock emission component, one gets an X-ray flare or X-ray
flattening, depending on the fall back accretion is steady or
not\footnote{So the shallow
decaying X-ray afterglow of some Swift GRBs may still be produced 
by the long activity of the central engine. In this model, the total 
energy needed to yield the X-ray
shallow-decay plateau should be much lower than 
that needed in the refreshed
forward shock model (Zhang et al. [2005] and the 
reference therein). Potential problems of our model are that how to get a
smooth translation from the shallow decay to the regular decay and why the 
eary forward shock emission in the X-ray band is so weak.}.

The dissipation stops abruptly after the reconnection events are
over. One then detects a steep decay component due to the
curvature effect.

Alternatively, the observed soft X-ray emission could also be due
to Comptonization of the mildly relativistic Alfv\'en turbulence
(excited in the wind by reconnection) off the photosphere photons
(e.g. Thompson 1994; M\'esz\'aros \& Rees 2000). If the Lorentz
factors of the intermittent outflow are highly variable, internal
shocks may still form if $\sigma$ is not very large. Significant
magnetic dissipation at the shock front is needed in order to get
a high radiation efficiency (Fan et al. 2004b).

\section{Linear polarization of the X-ray flares}
If X-ray flares are indeed powered by a Poynting-flux-dominated
jet, as argued above, a straightforward expectation is that the
detected emission should be linearly polarized. This is because
the magnetic fields from the central engine are likely frozen in
the expanding shells. The poloidal magnetic field component
decreases as $r^{-2}$, while the toroidal magnetic field component
decreases as $r^{-1}$. At the typical radius for ``internal''
energy dissipation, the frozen-in field is dominated by the
toroidal component. For an ultra-relativistic outflow, due to the
relativistic beaming effect, only the radiation from a very narrow
cone (with the half-opening angle $\leq 1/\Gamma$) around the line
of sight can be detected. As long as the line of sight is off the
symmetric axis of the toroidal magnetic field, the orientation of
the viewed magnetic field is nearly the same within the field of
view. The synchrotron emission from such an ordered magnetic field
therefore has a preferred polarization orientation (i.e. the
direction of the toroidal field). As a result, the linear
polarization of the synchrotron emission of each electrons can not
be significantly averaged out and the net emission should be highly
polarized (Lyutikov et al. 2003; Waxman 2003; Granot 2003). The
maximum polarization degree in an ordered field could be as high
as $\sim (60-70)\%$ (e.g., Lyutikov et al. 2003), but a lower
polarization degree is also expected since the dissipation
(through magnetic reconnections or internal shocks) process
may somewhat break the ordered field and lower the polarization degree
(e.g. Granot 2003). Assuming that in the radiation region the
strength ratio of the ordered field and the random field is $b$, the
detectable net polarization can be estimated as $\Pi_{\rm
net}\approx 0.6b^2/(1+b^2)$ (e.g. Granot 2003; Fan et al. 2004a).
It is hard to estimate $b$ without knowing the concrete energy
dissipation mechanism. In any case, a global ordered magnetic field
component should exist and usually plays an important role. In the
magnetic dissipation picture, the observed temporal variability does
not have to be related to internal shocks, which would potentially
destroy the ordered magnetic field configuration. Rather, the
variability is mainly related to the intermittent nature of the
accretion flow due to the interplay between the magnetic barrier and
gravity (Proga \& Begelman 2003). As a result, the ordered magnetic
fields generally survive in the dissipation regions.

Measuring polarization becomes a new direction in high energy
astronomy. New technologies are being invented, and many
polarimeter projects are under construction. In the X-ray band,
the ongoing projects include XPE (Elsner et al. 1997), SXRP
(Tomsick et al. 1997), PLEXAS (Marshell et al. 1998), POLAR
(Produit et al. 2005), etc. For example, the POLAR detector is
designed to have an energy range from a few keV up to several
hundred keV and a large field of view, which is very
suitable to detect X-ray flares following short GRBs. An important
issue is whether any of these detectors could perform a prompt
slew to the short GRBs localized by {\em Swift} (or other similar
GRB detectors). In some cases, weaker X-ray flares happen at an
even later epoch (e.g. $>10^4$ s for GRB 050724, Barthelmy et al.
2005). This somewhat eases the urgency of the prompt slew, but on
the other hand requires an even higher sensitivity. An ideal
instrument would be an XRT-like detector with polarization capability
on board a {\em Swift}-like GRB mission.

\section{Discussion}

We have argued that the X-ray flares detected following the short,
hard GRB 050724 should have been linearly polarized, if the progenitor
of this burst is a compact star merger, as is supported by its
association with an elliptical galaxy (Berger et al. 2005; Barthelmy et
al. 2005). The argument is achieved through gathering the X-ray flare
data and the insights from theoretical modeling of the GRB central
engines. The rapid decay following the flares suggest that the flares
are not related to afterglow emission. Rather, they reflect the
extended central engine activity. Based on the inferred mass accretion
rate ($\sim 0.01-0.001 ~{\rm M_\odot /s}$) from the merger scenarios,
the only mechanism to power the X-ray flares is the one involving
magnetic processes, and the jet should carry a dominant ordered
magnetic field component.
As a result, X-ray flares are expected to be linearly polarized.
Future X-ray polarimeters may be able to detect the polarized
signals from these flares.

Although throughout the Letter we are focusing on the X-ray flares
following short GRBs, the main argument may also apply to the
X-ray flares following long GRBs (Burrows et al. 2005; Piro et al.
2005), although the neutrino mechanism is not cleanly ruled out in
that case. Nonetheless, we suspect that those X-ray flares could
be polarized as well. 

\acknowledgments
We thank R. Perna for discussion and the referee for helpful
comments. This work is supported by NASA under grants NNG05GB67G,
NNG05GH92G, NNG05GH91G (for B.Z. \& Y.Z.F.), and NNG05GB68G (for
D.P.), as well as the National Natural Science Foundation (grants
10073022, 10225314 and 10233010) of China, and the National 973
Project on Fundamental Researches of China (NKBRSF G19990754) (for
Y.Z.F.).

\end{document}